\documentclass[twocolumn,pra]{revtex4}

\usepackage{latexsym}
\usepackage{graphics}
\usepackage{color}
\usepackage[latin1]{inputenc}
\usepackage{amssymb}
\usepackage{amsmath}
\usepackage{amsfonts}
\usepackage[dvips]{graphicx}

\usepackage{pdfpages}
\usepackage{epstopdf}

\begin{document}

\title{Differential phase extraction in dual interferometers exploiting the correlation between classical and quantum sensors}

\author{M. Langlois}
\author{R. Caldani}
\author{A. Trimeche}
\author{S. Merlet}
\author{F. Pereira dos Santos}
\email{franck.pereira@obspm.fr}
\affiliation{LNE-SYRTE, Observatoire de Paris, PSL Research University, CNRS, Sorbonne Universit\'es, UPMC Univ. Paris 06, \\ 61 Avenue de l'Observatoire, 75014 Paris, France}

\begin{abstract}

We perform the experimental demonstration of the method proposed in [Phys. Rev. A 91, 063615 (2015)] to extract the differential phase in dual atom interferometers. From a single magneto-optical trap, we generate two atomic sources, vertically separated and free-falling synchronously, with the help of an accelerated lattice. We drive simultaneous Raman interferometers onto the two sources, and use the correlation with the vibration signal measured by a seismometer to extract the phase of each interferometer. We demonstrate an optimal sensitivity of the extracted differential phase between the two interferometers, free from vibration noise and limited by detection noise, when the two interferometers are in phase.

\end{abstract}

% insert suggested PACS numbers in braces on next line
% insert suggested keywords - APS authors don't need to do this
%\keywords{}

%\maketitle must follow title, authors, abstract, \pacs, and \keywords
\maketitle
% body of paper here - Use proper section commands
% References should be done using the \cite, \ref, and \label commands

\section{Introduction}

Quantum sensors based on light pulse atom interferometry \cite{Borde1989}, such as gravimeters and gyrometers, have demonstrated high level of sensitivities and accuracies, comparable or better than conventional sensors \cite{Hu2013,Gillot2014,Freier2016,Dutta2016}. They find today applications in various fields, from fundamental physics to geophysics, and the transfer of this technology to the industry led in the last years to the development of commercial atomic gravimeters. The sensitivity of these sensors is limited in most cases by vibration noise, whose influence can be mitigated using passive isolation techniques \cite{LeGouet2008}, or auxiliary classical sensors for active isolation \cite{Hensley1999,Zhou2012,Hauth2013}, noise correction \cite{LeGouet2008,Merlet2009} or hybridization \cite{Lautier2014}. Nevertheless, when the measurand is derived from a differential measurement, performed on two interferometers interrogated at the same time, the vibration noise, which is then in common mode, can be efficiently suppressed. This technique has been used for the measurement of gravity gradients \cite{McGuirk2002,Sorrentino2014} and the precise determination of $G$ \cite{Fixler2007,Rosi2014}, for the measurement of rotation rates \cite{Gustavson1997,Canuel2006,Berg2015}, for the test of the universality of free fall with cold atoms \cite{Dimopoulos2007,Bonnin2013,Aguilera2014}. It is also of interest for the detection of gravitational waves \cite{Delva2006,Tino2007,Dimopoulos2008,Graham2013}. 

The differential phase, which is the phase difference between the two simultaneous interferometers, can be extracted simply from a fit to the ellipse obtained when plotting parametrically the output signals of the two interferometers \cite{Foster2002}. This method rejects the vibration noise efficiently but introduces in general large errors in the determination of the differential phase. Methods based on Bayesian statistics, which require an a priori knowledge of the phase noise of the interferometer, have been proven more accurate \cite{Stockton2007,Varoquaux2009,Chen2014}. Other methods, which use a simultaneous third measurement \cite{Rosi2015}, or direct extraction of the individual phases \cite{Bonnin2015}, also allow for the retrieval of the differential phase with negligible bias.

In this paper, we perform the experimental demonstration of the alternative method proposed in \cite{Pereira2015}. The correlation between the individual interferometer measurements and the vibration phase estimated from the measurement of an auxiliary seismometer allows us to recover the visibility of the interferometer fringes in the presence of large vibration phase noise and to extract the phase of each interferometer. The differential phase is then simply obtained by subtracting these two phases. This method of phase extraction, which was first demonstrated in \cite{Merlet2009} for a single interferometer, has been employed recently in \cite{Barrett2015} for simultaneous interferometers performed on two different atomic species. In the latter case, and by contrast with the situation we study here, the difference in the scale factors between the two interferometers reduces the correlation between the two extracted phases, degrading the rejection efficiency of the vibration noise. Here, we operate a dual atom interferometer on a single atomic species in a gradiometer configuration, with two interferometers separated along the vertical direction. Having the same scale factors, the better correlation between those two simultaneous interferometers enables us to reject more efficiently the common vibration noise. We demonstrate an optimal sensitivity in the differential phase extraction, limited by the detection noise, for in-phase operation of the two interferometers.

\section{Principle of the experiment}

\begin{figure}[ht]
        \centering
      \includegraphics[width=8.5 cm]{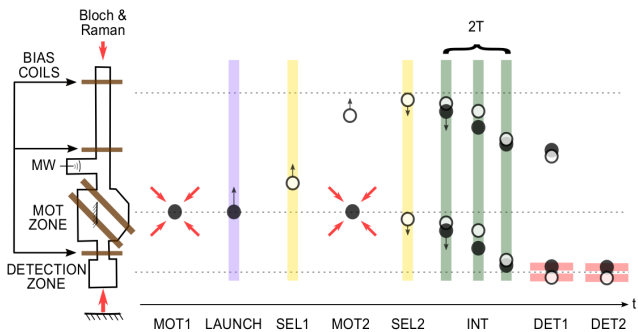}
\caption{Scheme of the experimental setup and measurement sequence.  Clouds displayed as open circles are in the state $|F=1\rangle$, clouds displayed as full circles in $|F=2\rangle$. MW: microwave antenna, SEL: selection, INT: interferometer, DET: detection.}
        \label{fig:expsequence}
   \end{figure} 
The experimental setup and the time chart of the measurement sequence are displayed in figure \ref{fig:expsequence}. We start by loading with a 2D magneto-optical trap (2D MOT) a 3D mirror MOT, realized with four independent beams, two of them being reflected by the surface of a mirror placed under vacuum. We trap about $3\times10^8$ $^{87}$Rb atoms within 480~ms, and cool them down to about $1.8~\mu$K with a far detuned molasses before releasing them from the cooling lasers in the $|F=2\rangle$ hyperfine ground state. Right after, they are launched upwards using a Bloch elevator \cite{Cadoret2008}. They are first loaded in a co-moving lattice, realized with two counterpropagating laser beams, whose intensity is adiabatically ramped up to a lattice depth of about $40~E_r$ within 200 $\mu$s, and whose acceleration follows the Earth gravity acceleration $g$. The lattice acceleration is then set to about $80g$ upward by ramping the frequency difference between the two lattice beams up to 4.5~MHz within 2.25~ms. The lattice intensity is then adiabatically switched off, leaving the atoms in free fall with an initial velocity of 1.76~m/s. The efficiency of the launch is about 50 \%. The launched atoms are then selected in the $|F=1, m_F=0\rangle$ state with a 0.8~ms long microwave pulse followed by a laser pushing pulse which removes the atoms remaining in $|F=2\rangle$ state. To lift the degeneracy between the different Zeeman sublevels, a bias field of 400~mG is applied. While these atoms are moving upwards, we load a second atomic cloud in the 3D mirror MOT for 100~ms. This second cloud is then cooled down to 1.8~$\mu$K and gets released from the molasses beams at the very moment when the first one reaches its apogee. We then apply the same selection sequence to prepare the second cloud in the $|F=1, m_F=0\rangle$ state. We arranged the second sequence so that the preparation of the second cloud does not affect significantly the first one: the first cloud being in the $|F=1\rangle$ state is merely perturbed by the scattered light from the MOT lasers (the repumping light in the second MOT is adjusted so as not to be saturated), and remarkably, for the launch velocity we use, the second microwave pulse, of 1.8~ms duration, drives a close to $2\pi$ pulse on the first atomic cloud. This is due to a favourable variation of the microwave coupling with vertical position. Most of the atoms of the first cloud thus remain in the $|F=1,m_F=0\rangle$ state.

At a delay of 32~ms after the release of the second cloud, we apply a sequence of three counterpropagating Raman pulses, equally separated in time, onto the two free falling atomic clouds. The Raman transitions couple the two hyperfine states $|F=1\rangle$ and $|F=2\rangle$ via a two photon transition, and impart a momentum transfer $\hbar k_{eff}$ to the atoms. $k_{eff}=k_1-k_2$ is the effective wavevector of the Raman transition, with $k_1$ and $k_2$ the wavevectors of the two Raman lasers. The pulse sequence allows to separate, deflect and recombine the atomic wavepackets, creating simultaneously two vertically separated Mach Zehnder atom interferometers. The atomic phase-shift at the output of each interferometer is then given by: $\Delta\Phi=\phi_{1}-2\phi_{2}+\phi_{3}$, where $\phi_{\mathrm{i}}$ is the phase difference between the two counterpropagating Raman lasers at the position of the atoms at the i-th Raman pulse. The Raman beams are vertically aligned, which makes these interferometers sensitive to the local gravity accelerations \cite{Kasevich1991}. The interferometer phase shifts are given by $k_{eff} g_1 T^2$ and $k_{eff} g_2 T^2$, where $g_1$ and $g_2$ are the gravity accelerations at the altitudes of the two clouds, where $T$ is the time separation between consecutive Raman pulses. The duration of the $\pi$ Raman pulse is 8~$\mu$s, which corresponds to a Rabi frequency of 62.5~kHz. The size of the vacuum chamber limits the maximum duration of the interferometers $2T$ to about 160~ms. After the interferometer sequence, the two atomic clouds are detected one after the other by fluorescence, using a state selective detection which measures the populations in each of the two output ports of the interferometers, corresponding to the two hyperfine states $|F=1\rangle$ and $|F=2\rangle$.

\begin{figure}[ht]
        \centering
      \includegraphics[width=8 cm]{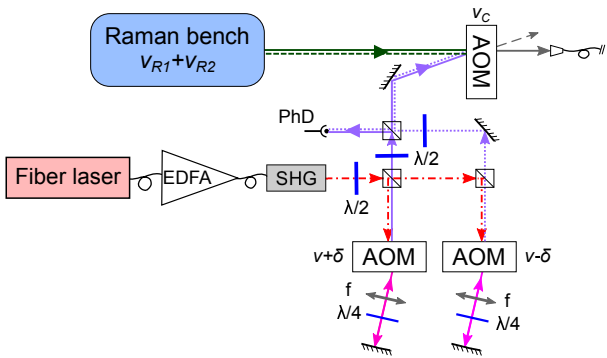}
\caption{Scheme of the lattice beams laser setup. The lattice beams generated by a frequency doubled fiber telecom laser are combined with the Raman beams using an AOM.}
        \label{fig:setup}
   \end{figure}

The laser system we use for cooling, detecting and driving the interferometer pulses is based on semiconductor laser sources, and is described in detail in \cite{Merlet2014}. For the Bloch elevator, we generate a lattice laser using frequency doubling techniques (see figure \ref{fig:setup}). A DFB diode laser at 1560~nm is first amplified by a fiber amplifier up to 5~W, and seeds a high power resonant frequency doubling module (from the company Muquans) which delivers up to 3~W at 780~nm. The frequency of the seed is adjusted such that the frequency doubled light is blue detuned from the $^{87}Rb$ D2 transition by 50~GHz. The output of the doubler is then split into two beams, each one being frequency shifted with a double pass acousto-optic modulator (AOM) before being recombined with orthogonal linear polarisations using a polarizing beam splitter cube. The combined beam is diffracted by a last AOM, which is used for switching and controlling the intensity of the lattice beams. The Raman beams, which are derived from the first laser system, have the same linear polarisation. They are overlapped with the lattice beams using the very same AOM, into which they are injected at an angle, along the direction of the first diffracted order. This arrangement allows to overlap the diffracted lattice beams with the non-diffracted Raman beams in order to inject them into the same fiber, and takes advantage of the fact that the two beams are not used at the same time. As for the switching of the Raman beams, it is performed with the combination of an AOM and a mechanical shutter (an optical scanner), both located in the first laser system. 

\begin{figure}[ht]
        \centering
      \includegraphics[width=8 cm]{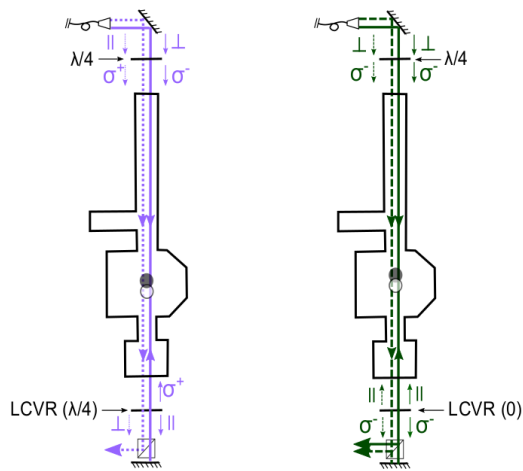}
\caption{Polarization configuration of lattice (left) and Raman (right) beams in the vacuum chamber. The combination of a quarter-wave plate and an adjustable retarding plate (LCVR) with a polarizing beamsplitter cube allows to obtain the required polarizations for realizing an accelerated lattice along one direction only and for driving counterpropagating Raman transitions.}
        \label{fig:setup2}
   \end{figure}

Raman and lattice beams are injected into a common polarization maintaining fiber, out of which we get a total power of 500~mW for the lattice beams (for a doubler output of 2~W), and 26~mW for the Raman beams. The ratio between the intensities of the two Raman beams is adjusted to cancel the differential light shift. The beams get collimated to a $1/e^2$ radius of 3.75~mm and enter the vacuum chamber through the top of the experiment. The polarization configuration for these beams is displayed in figure \ref{fig:setup2}.

A fixed quarter wave plate converts the linear polarisations into circular ones : $\sigma_+ - \sigma_-$ or $\sigma_- - \sigma_-$ for the lattice or Raman beams, propagating downwards. At the bottom of the chamber, the beams pass through a liquid crystal variable retarder plate (LCVR), a polarizing beamsplitter cube (PBC) with proper axes at $45^{\circ}$ with respect to the axes of the LCVR and finally get retroreflected onto a mirror. The retardance of the LCVR plate is set to $\lambda/4$ during the lattice launch, such that one of the two downward lattice beams gets trashed by the PBC. Without the cube, this retroreflecting geometry would result in two lattices accelerated in opposite directions. During the interferometer phase, the retardance is set to zero, so that the Raman upward beams are linearly polarized. This polarization arrangement allows to drive $\sigma_- - \sigma_-$ transitions between the two $|F=1, m_F=0\rangle$ and $|F=2, m_F=0\rangle$ states. As a side effect, $\sigma_- - \sigma_+$ transitions are allowed, which couple $|F=1, m_F=0\rangle$ to $|F=2, m_F=-2\rangle$ or $|F=2, m_F=+2\rangle$ depending on the direction of the Raman effective wavevector. This forces us to operate the interferometers with large bias fields of hundreds of mG, for which these parasitic transitions are non resonant.   

With such a retroreflected geometry, the phase difference between the Raman lasers is linked to the position of the mirror. Without isolation, fluctuations of this position due to ground vibrations can induce significant interferometer phase noise, washing out the interferometer fringes, in the urban environment of the center of Paris. This vibration noise is recorded with a low noise seismometer (Guralp 40T), placed next to the mirror, during the interferometer sequence. This allows to calculate an estimate of the common mode phase shift induced by parasitic vibrations onto the two interferometers. The correlation between the classical signal provided by the seismometer and the phase shifts of the quantum sensor can be exploited to post-correct the atomic measurement \cite{LeGouet2008}, to recover the interferometer fringes in the presence of a large noise \cite{LeGouet2008,Merlet2009}, or to correct the interferometer phase in real-time in order to keep the interferometer operating at mid-fringe where its sensitivity is maximal \cite{Lautier2014}. Given that our measurements are performed without any vibration isolation, the vibration noise quickly dominates over all sources of interferometer phase noise and can amount to several radians, even for the relatively short interferometer times we use here ($2T$ is at most equal to 120~ms).   

\section{Results}

\begin{figure*}
        \centering
      \includegraphics[width=18 cm]{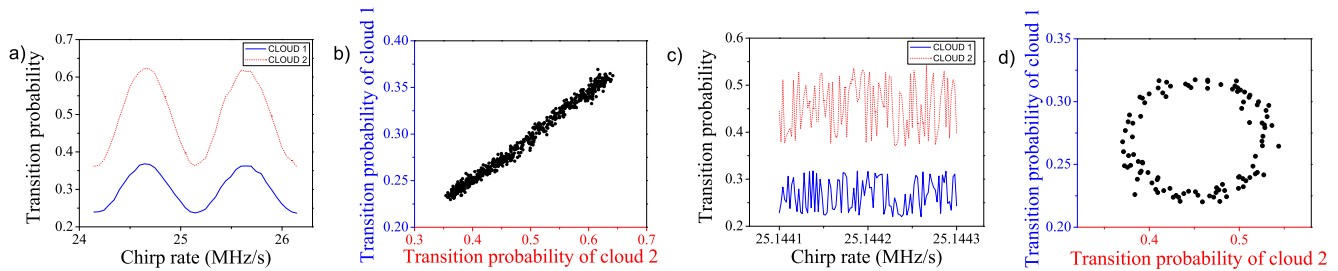}
\caption{Interferometer fringe patterns for the two simultaneous interferometers, for two different interferometer times $2T=2$~ms (a) and $2T=70$~ms (c). The interferometer phases are scanned by varying the frequency chirp applied to the Raman lasers. The parametric plot of the transition probabilities for in phase (b) (resp. out of phase (d)) interferometers gives a line (resp. an ellipse).}
        \label{fig:phase1}
   \end{figure*}

We start by illustrating the effect of the vibration noise onto the interferometer signal. Figure \ref{fig:phase1} displays the fringes recorded for two different interferometer times $2T=2$~ms and $2T=70$~ms. Here, the interferometer phase is scanned by varying the frequency chirp one needs to apply to the Raman laser frequency difference in order to compensate for the increasing Doppler effect and keep the Raman transitions resonant at each pulse. While the fringes are clearly visible for $2T=2$~ms, they are washed out by the vibration noise for $2T=70$~ms. The contrast of the fringes is significantly better for the second cloud than for the first one, and gets lower for larger $T$. This loss of contrast results mostly from the expansion of the cloud and from its convolution with the transverse intensity profile of the Raman lasers: coupling inhomogeneities get larger for larger atomic cloud sizes, and are thus larger for the launched cloud which has been expanding for much longer times compared to the second cloud. In the presence of large vibration noise, the phase fluctuations of the two interferometers are strongly correlated. This is evidenced by plotting the transition probabilities parametrically, which, in general and in particular here for $2T=70$~ms, gives an ellipse (as illustrated in figure \ref{fig:phase1} (d)). Yet, when the two interferometers are in phase, the parametric plot gives a straight line as displayed in figure \ref{fig:phase1} (b). A direct fit of the ellipse gives access to the differential phase, but the adjustment is in general biased, except when the differential phase is exactly $\pi/2$. In addition, such a fit cannot retrieve a null differential phase. 
In our experiment, the differential phase between the two interferometers can easily be tuned by changing the amplitude of the bias magnetic field, as both interferometers have large, but different, phase shifts due to gradients of the applied magnetic field. Indeed, the magnetic field profiles are different across the atom trajectories of the two interferometers. It varies for the first cloud from about 400 mG to 200 mG in between the first and last pulse of the interferometer, and for the second cloud oppositely from 200 mG to 400 mG.  

Instead of the ellipse fitting method, we use here another method for the extraction of the differential phase. Knowing the well calibrated scale factor of the seismometer and the transfer function of the interferometers versus acceleration noise \cite{Cheinet2008}, we calculate out of its signal an estimate of the common mode vibration phase shift experienced by the two interferometers. We plot the measured transition probabilities of the two interferometers as a function of this vibration induced phase shifts. 

\begin{figure}[ht]
        \centering
      \includegraphics[width=8 cm]{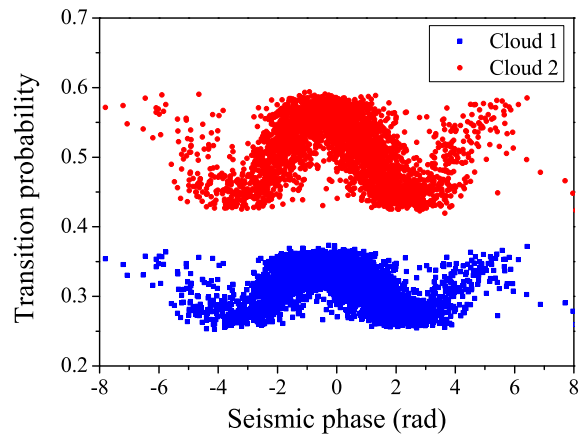}
\caption{Measured transition probabilities versus vibration phase shift, estimated from the simultaneous measurement of a low noise seismometer. The total interferometer time is $2T=120$~ms. The retrieval of the fringe pattern reveals the correlation between the interferometer phase fluctuations and the vibration noise recorded by the classical sensor.}
        \label{fig:phasevib}
   \end{figure}

Figure \ref{fig:phasevib} displays such plots, for given values of the bias magnetic fields and $2T=120$~ms. Here, visible fringe patterns are recovered, with the interferometer phase being randomly scanned by the vibration noise. We can then fit each of these fringe pattern using the following formula: 
\begin{equation}
P_i=A_i+\frac{C_i}{2} \textrm{cos}(D_i \times \Phi_{vib,S} + \Phi_i)
\end{equation}
with $A_i$ the offset, $C_i$ the contrast, $\Phi_{vib,S}$ the calculated vibration induced phase shift and $\Phi_i$ the phase shift of the i-th interferometer. $D_i$ is a coefficient which accounts for an eventual mismatch in the calibration of the seismometer. In practice, we find that $D_1$ and $D_2$ deviate from 1 by a few percents, due to the non flat response function of the seismometer \cite{LeGouet2008}. Such fits allow to extract the individual phases of the two interferometers, from which the differential phase is calculated. 

\begin{figure}[ht]
        \centering
      \includegraphics[width=8 cm]{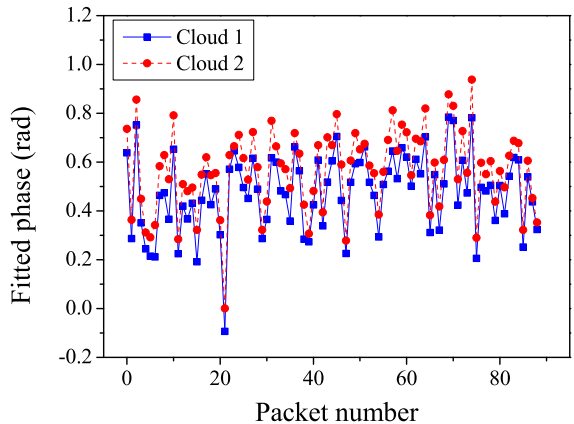}
\caption{Interferometer phases extracted from consecutive fits of the fringes for subsets of 40 measurements. Total interferometer time $2T=120$~ms.}
        \label{fig:phases}
   \end{figure}

To assess the sensitivity of this method, we divide a series of about 4000 consecutive measurements into packets of 40 measurements, which we individually fit. The phases extracted from these fits are displayed on figure \ref{fig:phases} for the two clouds. For these measurements, one can note that the fluctuations of the extracted phases, of order of about 0.6 rad peak-to-peak, are correlated for the two clouds. 

\begin{figure}[ht]
        \centering
      \includegraphics[width=8 cm]{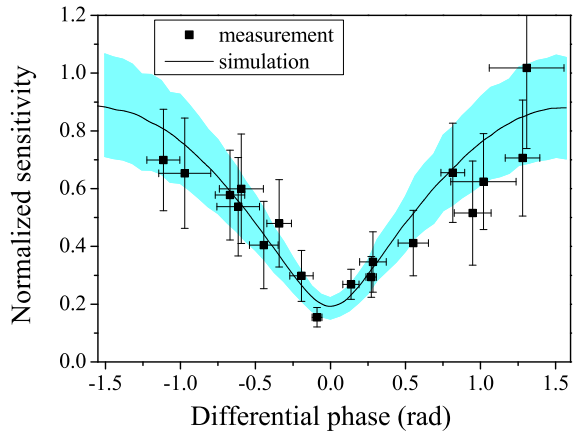}
\caption{Sensitivity of the differential phase extraction as a function of the extracted differential phase. The points represent the results of the measurements. The line corresponds to the result of the numerical simulation, and the shaded area to its uncertainty for a finite number of 4000 data samples.}
        \label{fig:rejection}
   \end{figure}

We then calculate the Allan standard deviation (ASD) of the fitted phase fluctuations of the two interferometers, and of their difference. 
\begin{figure}[ht]
        \centering
      \includegraphics[width=8 cm]{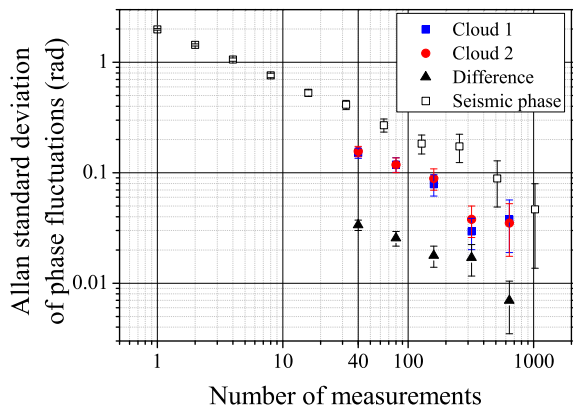}
\caption{Allan standard deviations of the vibration phase noise (seismic phase displayed as open squares), the individual retrieved interferometer phases (full squares and circles) and the differential phase (diamonds). Total interferometer time: $2T=120$~ms.}
        \label{fig:asds}
   \end{figure} 
Figure \ref{fig:asds} displays such ASDs for in phase interferometers with $2T=120$~ms. The ASDs average as white noise, and correspond to phase sensitivities of 150~mrad/packet of 40 shots, or equivalently of 1 rad/shot, for the individual interferometers. In comparison, the ASD of the induced vibration noise of 2 rad/shot. The gain is not significant, due to the poor quality of the correlation, as can be seen in figure \ref{fig:phasevib}. Despite this, the ASD of the differential phase is significantly less, of about 33~mrad/packet (or equivalently of 208~mrad/shot), and lies not far from the limit set by our detection noise (120~mrad/shot on the differential measurement). This puts into evidence the existence of strong correlations between the values of the fitted phases for the two interferometers, which are suppressed when taking their difference. 
From \cite{Pereira2015}, we expect this correlation, and thus the rejection of the common mode vibration noise, to decrease when the differential phase increases. We have investigated this loss of sensitivity by repeating this analysis for different differential phases in a range from -1.5 to 1.5~rad (this range corresponds to a variation of the current in the bias coils of 10\%). Figure \ref{fig:rejection} displays the sensitivity of the differential phase extraction we obtain as a function of the differential phase. In order to highlight the effect of this correlation, this sensitivity is normalized by the one we would expect in the absence of any correlation between the two interferometer phases, which correspond to the quadratic sum of the sensitivities $\sigma_i$ obtained individually: $\sqrt{\sigma_1^2+\sigma_2^2}$. We indeed observe that the sensitivity reaches its best level for in phase interferometers. 

The results are compared with numerical simulations, which we take as representative as possible of the experiment. In these simulations, we generate the transition probabilities of the two interferometers, by randomly drawing the vibration phase estimated by the seismometer $\Phi_{vib,S}$ in a Gaussian distribution. We use for the contrasts of the two interferometers the average values of the fitted contrasts, 10\% and 6\%. We add to the transition probabilities uncorrelated Gaussian detection noises with standard deviation $\sigma_P=3\times10^{-3}$, equal to the measured detection noise. We also randomly draw $\delta \Phi_{vib}$ the difference between the vibration phase $\Phi_{vib}$ and its estimate $\Phi_{vib,S}$ in a Gaussian distribution. The standard deviation of $\delta \Phi_{vib}$ is adjusted so as to obtain the same sensitivity of 1~rad/shot as in the measurements when extracting the individual phases from the simulated data using the method described above. This adjustment corresponds to a standard deviation of $\delta \Phi_{vib}$ of 620~mrad/shot. We then repeat the simulations for various differential phases and the normalized sensitivity we obtain in the simulation for the extraction of the differential phase is displayed on figure \ref{fig:rejection} as a line. The shaded area corresponds to the uncertainty in the estimation of the Allan standard deviations, given that the number of data samples in the measurements is finite. This confidence area is estimated from the dispersion of the results obtained when repeating numerical simulations with different sets of random draws and with the same number of data samples as for the measurements (typically 4000 measured samples for each differential phase, to be compared with the 500 000 draws generated to calculate the normalized sensitivity displayed as a line). We find a perfect agreement between the experimental results and the corresponding simulation, given that most of our measurements lie in the shaded confidence area and that the uncertainties in the measured sensitivities match the width of the simulated confidence area.

\begin{figure}[ht]
        \centering
      \includegraphics[width=8 cm]{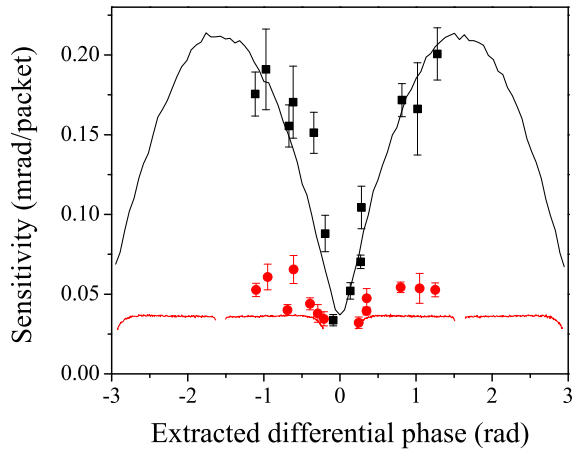}
\caption{Comparison with the ellipse fitting technique. Points display the results of the measurements (squares: our method, circles: ellipse fitting), and lines of the simulation (black line: our method, red (light gray) line: ellipse fitting).}
        \label{fig:fitellipse}
   \end{figure}

We finally compare the technique studied here with the direct fit of the ellipse, using the same fitting procedure as \cite{Foster2002}, both in the measurements and the simulations. Figure \ref{fig:fitellipse} displays the sensitivities of the differential phase (not normalized here) obtained with these measurements, displayed as points, and with the simulations, displayed as lines. For a quantitative match between experiments and simulation, we had for this simulation to decrease the amplitude of the vibration noise measured by the seismometer $\sigma_{\Phi_{vib,S}}$ from 2 to 1.8 radian (the latter value corresponding to an average amplitude over all the measurements of the measured seismic noise, which fluctuates from one measurement to the other) and to increase the amplitude of $ \sigma_{\delta \Phi_{vib}}$ to 0.86 rad. 

With respect to the method presented here, the ellipse fitting technique rejects better the common vibration noise. Its sensitivity does not depend on the value of the differential phase, and equals the optimal sensitivity we obtain with our method for a null differential phase. On the other hand, the simulations show that the ellipse fitting technique leads to a biased differential phase, whereas our method is in principle unbiased. Also, by contrast with our method, the ellipse fitting routine cannot extract differential phases close to zero \cite{Foster2002}, and suffers from ambiguities in the determination of the differential phase, which complicates the extraction, especially close to $\pi/2$. This explains the discontinuities of the red line in the figure \ref{fig:fitellipse}. Finally, one can note significant deviations of the measurements from the simulation for the ellipse fitting. We attribute these mismatches to variations from one measurement to the other of the detection noise, due to changes in the contrast of the interferometers and in the number of detected atoms. 

Assessing experimentally our ability of extracting the differential phase accurately, as claimed in \cite{Pereira2015}, requires a method for varying this differential phase in an accurate way. This cannot easily be realised with magnetic field gradient phase shifts. An alternative method consists in changing the frequencies of the Raman lasers at the second pulse, such as demonstrated in \cite{Biedermann2015}.

\section{Conclusion}

We have performed the experimental validation of the method proposed in \cite{Pereira2015} for extracting the differential phase in dual atom interferometers. The experiment was performed on an atom gradiometer setup, consisting in two simultaneous atom gravimeters separated along the vertical direction. We have exploited the correlations between the individual noisy measurements of each interferometer and the estimates of the phase noise introduced by parasitic ground vibrations to determine the individual phases of each interferometer, out of which the differential phase is straight-forwardly obtained. We find that the sensitivity of the differential phase extraction is optimal, and close to the limit set by the detection noise, when the two interferometers are in phase. We have finally briefly compared this method with the simple and more often used ellipse-fitting method. A thorough comparison with other techniques would be of interest, but lie beyond the scope of this paper. In the future, we will demonstrate the accuracy of this method, thanks to the fine tuning of the differential phase obtained by changing the frequency of the Raman lasers at the central $\pi$ pulse. As shown in \cite{Roura2017} and pointed out in \cite{Rosi2017}, with a specific adjustement of this frequency change and thus of the corresponding Raman wavevector, the gradiometer differential phase can be compensated, which allows for a precise determination of the gravity gradient independently of the gradiometric baseline. Our method for extracting the differential phase thus appears perfectly suited to the implementation of this compensation technique since it works best for a null differential phase.   
 
\section{Acknowledgements}

This work was supported by CNES (R\&T R-S15/SU-0001-048), by DGA (Gradiom project), by the ``Domaine d'Int{\'e}r{\^e}t Majeur'' NanoK of the R{\'e}gion Ile-de-France, by the CNRS program ``Gravitation, R{\'e}f{\'e}rences, Astronomie, M{\'e}trologie'' (PN-GRAM) co-funded by CNES and by the LABEX Cluster of Excellence FIRST-TF (ANR-10-LABX-48-01), within the Program ``Investissements d'Avenir'' operated by the French National Research Agency (ANR). M.L. thanks Muquans for financial support.

M.L. and R.C. contributed equally to this work.

\end{document}